\documentstyle[12pt]{article}
\newcommand{\be}{\begin{equation}}
\newcommand{\ee}{\end{equation}}
\newcommand{\bea}{\begin{eqnarray}}
\newcommand{\eea}{\end{eqnarray}}
\newcommand{\nn}{\nonumber}
\newcommand{\p}{\phi}
\newcommand{\q}{\omega}
\newcommand{\de}{\delta}
\newcommand{\hp}{{\hat \phi}}
\begin{document}

\title{Non-Abelianizable First Class Constraints}
\author{Farhang Loran\thanks{e-mail:
loran@cc.iut.ac.ir}\\ \\
  {\it Department of  Physics, Isfahan University of Technology (IUT)}\\
{\it Isfahan,  Iran}}
\date{}
\maketitle

\begin{abstract}
We study the necessary and sufficient conditions on Abelianizable
first class constraints. The necessary condition is derived from
topological considerations on the structure of gauge group. The
sufficient condition is obtained by applying the  theorem of
implicit function in calculus and studying the local structure of
gauge orbits. Since the sufficient condition is necessary for
existence of proper gauge fixing conditions, we conclude that in
the case of a finite set of non-Abelianizable first class
constraints, the Faddeev-Popov determinant is vanishing for any
choice of subsidiary constraints. This result is explicitly
examined for SO(3) gauge invariant model.
\end{abstract}
\newpage

\section{Introduction}
 A gauge theory, in general, possesses a set of first class
 constraints, $\p_i$, $i=1,\cdots,N$, satisfying the algebra
 \be
 \{\p_i,\p_j\}=\sum_{k=1}^N U_{ij}^{\ \ k}\p_k,\hspace{1cm}i,j=1,\cdots,N,
 \label{int1}
 \ee
 in which the structure functions $U_{ij}^{\ \ k}$'s are generally
 some functions of phase space coordinates. One of the most
 interesting questions in constraint systems is the possibility of converting
 a given set of first class constraints
 to an equivalent Abelian set. By definition, Abelian constraints
 commute with each other, i.e. their Poisson brackets with each
 other are vanishing identically. There are various motivations for examining
 such a possibility.
 For example, first class constraints are generators of gauge
 transformation: $\de_i^\p F(z)=\{F(z),\p_i\}$ \cite{Dirac}.
 Since $U_{ij}^{\ \ k}$ are functions of phase space coordinates, the
 full generator of gauge transformation is a nontrivial combination of
 first class constraints \cite{Pons}. This combination is the simplest
 if first class constraints are Abelian i.e. when the Poisson brackets of
 these constraints with each other vanish identically.
 \par
 Abelianization of first class constraints can also result in two
more important simplifications. First, following Dirac's
arguments, quantization of a set of first class constraints
satisfying the algebra (\ref{int1}), where $U_{ij}^{\ \ k}$'s are
not $c$-numbers, requires a definite operator ordering
\cite{Dirac}. That is because  in Dirac quantization, physical
states are
 defined as null eigenstates of the operators $\hp_i$'s,
 \be
 \hp_i\left|\mbox{phys}\right>=0,
 \label{int2}\ee
 in which the operator $\hp_i$'s are defined corresponding to the
 constraints $\p_i$'s. Definition (\ref{int2}) and the algebra (\ref{int1})
 are consistent if the operators $\hat{U}_{ij}^{\ \ k}$'s, defined
 corresponding to the structure functions $U_{ij}^{\ \ k}$'s, sit on
 the left of the operators $\hp_i$'s similar to
 Eq.(\ref{int1}). The existence of such an operator ordering is not
 evident generally. Apparently, when first class constraints are Abelian,
 no such operator ordering
 should be considered. Second, in BRST formalism, the algebra (\ref{int1}),
 in general, leads to a complicated expansion of the BRST charge in terms
 of the ghosts.
 When first class constraints are Abelian, the generator of BRST
 transformation can be
 recognized in the most simple way \cite{Hen}.
 \par
 Different methods for Abelianization of first class constraints
 are studied. Example are, Abelianization via constraint resolution
 \cite{Hen,Henbook,Gold} or via
generalized canonical transformation for general non-Abelian
constraints (that satisfy a closed algebra) \cite{Berg}. In
reference \cite{Gog} the authors study Abelianization via Dirac's
transformation. In this method, one assumes that linear
combinations of non-Abelian first class constraints (satisfying a
closed algebra) exist that converts the given set of non-Abelian
constraints to an equivalent set of Abelian constraints. In this
way the problem of Abelianization is led to that of solving a
certain system of first order linear differential equations for
 the coefficients of these linear combinations.
 In \cite{Abelian}, it is  shown  that mapping each first
 class constraint to the surface of the other constraints, results
 in  Abelian first class constraints. In \cite{Second} it is shown
 that the maximal Abelian subset of second class constraints can be
 obtained in the same way.
 \par
 The domain of validity and/or applicability of the above methods
 can be determined by studying the necessary conditions on
 Abelianizable first class constraints. Topology
 of gauge group at each point $p$ of the phase space, which is uniquely
 determined by the structure coefficients $U_{ij}^{\ \ k}(p)$,
 provides the necessary tools for this purpose. In fact, if at some
 point $p$, a non-Abelian set of first class constraints can be
 made Abelian, the corresponding gauge group should
 be topologically equivalent to the Abelian gauge groups, i.e.
 the group of Euclidean translations.
 \par
 In \cite{Henbook} a method for Abelianization of first class
 constraints is proposed, which is based on the theorem of implicit
 differentiation (or the theorem of implicit function) and gives a
 sufficient condition on Abelianizable first class constraints.
 According to that theorem,
 if at some point $p$, $d\p$ is maximal (see section 3), one can
 in principle, solve the equations $\p_i(z_1,\cdots,z_N;z'_a)=0$,
 $i=1,\cdots,N$ as $z_i=z_i(z'_a)$, $i=1,\cdots,N$. It is shown in
 \cite{Henbook} that the constraints $\psi_i=z_i-z_i(z'_a)$, are
 Abelian. Therefore, maximality is the sufficient condition for
 constraints to be Abelianizable. Using these results, we conclude
 that $U_{ij}^{\ \ k}(p)$'s determine whether the maximality condition is
 satisfied at $p$ or not.
 \par
 Violation of maximality causes serious problems. For example, the norm of
 the constraint surface is not well defined in the neighborhood of
 maximality-violated regions. Furthermore, the necessary condition on
 subsidiary constraints (gauge fixing conditions) $\q_i$'s, i.e.
 $\det(\{\p_i,\q_j\})_p\neq 0$,
 can not be satisfied if maximality is violated at $p$. This means that
 Faddeev-Popov determinant is vanishing regardless our choice of subsidiary
 constraints.
 The equivalence of Lagrangian and
 Hamiltonian formalism for constraint systems is also proven under the
 assumption that primary constraints satisfy the maximality condition all
 over the phase space \cite{Battle}.
 \par
 Of course it should be noted that here we study only
 constraint systems with a finite set of first class constraints and it is not
 straightforward to generalize the results of this paper to systems with an infinite set
 of constraints, e.g. the SU(N) Yang-Mills theory (See appendix 2).
 \par
 The organization of paper is as follows. In section 2, we study
 the topological conditions on Abelianizable first class
 constraints. In section 3, we discuss the maximality condition as the sufficient condition
 in two instructive ways: in subsection 3.1 by reviewing the Abelianization method of
 ref.\cite{Henbook} and in subsection 3.2 by studying the local structure of gauge orbits.
 In section 4, we examine our results explicitly by considering two simple
 examples. Section 5, is devoted to summary and conclusion. There
 are also two short appendices. Appendix 1 is a review of the
 method of obtaining orthogonal bases of a given vector space. The
 second appendix is a review of the constraint algebra of $SU(N)$
 Yang-Mills model.
\section{Topological Considerations, A Necessary Condition}
 Assume a finite set of irreducible first class constraints $\p_i$,
 $i=1,\cdots,N$. By definition,
 \be
 \{\p_i,\p_j\}=U_{ij}^{\ \ k}\p_k,
 \label{a1}
 \ee
 where $U_{ij}^{\ \ k}(z^\mu)$ are some functions of phase space
 coordinates $z^\mu$'s. $\{\p_i,\p_j\}$ stands for the Poisson bracket of
 $\p_i$ and $\p_j$ defined as follows:
 \be
 \{\p_i,\p_j\}=\frac{\partial \p_i}{\partial
 z^\mu}J^{\mu\nu}\frac{\partial \p_j}{\partial z^\nu},
 \ee
 where $J^{\mu\nu}=\{z^\mu,z^\nu\}$ is a full rank antisymmetric tensor,
 e.g the symplectic two form,
 \be
 J=\left(\begin{array}{cc}0&1\\-1&0\end{array}\right)
 \label{symp}
 \ee
 The gauge transformation of any function of phase space, $F(z)$
 is given by $\de^\p_i F(z^\mu)=\{F,\p_i\}|_\Phi$ where  $\Phi$ is
 the constraint surface corresponding to the
 constraints $\p_i=0$, $i=1,\cdots,N$ \cite{Dirac}.
 Using Eq.(\ref{a1}), and the Jaccobi identity for Poisson brackets,
 one can show that for an arbitrary analytic function of phase
 space coordinates $F(z)$,
 \begin{eqnarray}
 [\de^\p_i,\de^\p_j]F(z)&=&
 \left\{\{F(z),\p_j\},\p_i\right\}|_{\Phi}-
 \left\{\{F(z),\p_i\},\p_j\right\}|_{\Phi}\nonumber\\
 &=&\left\{\{\p_i,\p_j\},F(z)\right\}|_{\Phi}\nonumber\\
 &=&\left\{U_{ij}^{\ \ k}\p_k,F(z)\right\}|_{\Phi}\nonumber\\
 &=&-U_{ij}^{\ \ k}\{F(z),\p_k\}|_{\Phi}\nonumber\\
 &=&-U_{ij}^{\ \ k}\de^\p_kF(z).
  \end{eqnarray}
 Consequently, $\de^\p_i$'s are elements of a Lie algebra with
 structure coefficients $-U_{ij}^{\ \ k}$'s:
 \be
 [\de^\p_i(p),\de^\p_j(p)]=-U_{ij}^{\ \ k}(p)\de^\p_k(p),
 \hspace{1cm}p\in\Phi,
 \ee
 The corresponding Lie group
 is called the gauge group and the gauge orbits are the integral curves of
 $\de^\p_i(p)$'s.
 \par
 The main concept in our study is the concept of equivalence
 of two sets of constraints.
 Two sets of constraints are said to be equivalent at some point $p$ of the
 phase space if 1) the corresponding constraint surfaces are similar
 at some neighborhood of $p$ and 2) the resulting gauge
 transformations are equivalent  \cite{Henbook,Govbook}.
 These conditions are fulfilled  using the  following definition of
 equivalence:
 \par
 {\bf Definition} The set of constraints $\psi_i$, $i=1,\cdots, N$
 are equivalent to $\p_i$'s, $i=1,\cdots,N$ at $p$ if
 1) $p\in\Phi$ and $p\in\Psi$, 2)
 $T_p\Phi$ is homeomorphic  to $T_p\Psi$ and 3) $G_p^\p=G_p^\psi$.
 \par
 $\Phi$($\Psi$) is the constraint surface corresponding to the constraints
 $\p_i$'s($\psi_i$'s) and $T_p\Phi$$(T_p\Psi)$ is the tangent space of
 $\Phi$$(\Psi)$ at $p$. $G^\p_p$ and $G^\psi_p$ are the gauge groups
 generated by $\de^\p_i(p)$'s and $\de^\psi_i(p)$'s respectively.
 Two topological spaces are said to be
 homeomorphic if there exist an invertible continuous
 map  between them \cite{Munk}. Since $\de_i^\p(p)$'s ($\de_i^\psi(p)$)
 expand some subspace of the tangent space $T_p\Phi$($T_p\Psi$) (see subsection 3.2) and the
 equivalence of two Lie groups $G_p^\p$ and $G_p^\psi$ requires that these
 subspaces be homeomorphic,\footnote{Equality, $=$, is a trivial
 homeomorphism given by the identity map} the second and third conditions in the
 definition given above are consistent with each other.
 \par
 If two topological spaces are homeomorphic, their topological invariants
 should be the same. It is important to note that this is a necessary and
 not a sufficient condition.  Example of topological invariants are
 connectedness and compactness \cite{Munk}.
  \par
 Abelianization of a set of non-Abelian first class constraints
 amounts to obtaining an equivalent set of Abelian constraints.
 The gauge group of Abelian constraints is
 homeomorphic  to the group of Euclidean translations, i.e. ${\cal R}^N$,
 where $N$ is the number of first class constraints.\footnote{Some directions
 of  ${\cal R}^N$ can be compactified. A compactified direction corresponds to a
 $U(1)$ gauge symmetry. The simplest example of such systems is the Friedberg model
 \cite{Fried}. Although we do not consider such cases here but
 generalization of the final result to these systems is straightforward.}
 These are simply-connected and non-compact spaces.  Consequently,
 the necessary (not sufficient) condition on constraints $\p_i$'s
 to be Abelianizable at some point $p$ is that the corresponding
 gauge group determined by $U_{ij}^{\ \ k}(p)$'s should be
 simply-connected and non-compact. For example, if $U_{ij}^{\ \ k}(p)$
 are the structure coefficients of some compact group e.g. $SO(N)$,
 then the corresponding constraints can not be made Abelian at
 $p$. We call such sets of first class constraints,
 non-Abelianizable constraints.
 \par
 As an example consider the $SO(3)$ gauge invariant model \cite{Govpap}
 where first class constraints are $L_i=\epsilon_{ijk}x_jp_k$, $i=1,2,3$ satisfying
 the algebra $\{L_i,L_j\}=\epsilon_{ijk}L_k$, in which
 $\epsilon_{ijk}$ is the Levi-Civita tensor. Consequently
 the  gauge group is compact (homeomorphic to ${\cal S}^2$)
 and $L_i$'s are non-Abelianizable.
 \section{Maximality, A Sufficient Condition}
 In this section we obtain the sufficient condition, which we call
 the maximality condition,
 on constraints to be Abelianizable. Obviously the sufficient condition is not satisfied
 in the case of non-Abelianizable constraints. As we will show this means that
 the Faddeev-Popov determinant in such systems is vanishing
 for any choice of subsidiary constraints. In the
 following two subsections, we follow two different
 methods to study the maximality condition. The first one is based on the
 Abelianization via constraint resolution which is a well known method \cite{Henbook}.
 In the second method we study the local structure of gauge orbits.
 Although the following methods are basically equivalent but they clarify different aspects
 of the maximality condition.
 \subsection{Resolution of Constraints}
 Here we review the Abelianization method introduced in
 \cite{Henbook}. Assume a set of first class constraints $\p_i$,
 $i=1,\cdots, N$ and the corresponding constraint surface $\Phi$.
 Consider a point $p\in \Phi$, where $d\p$ is maximal. Maximality
 here means that there exist a subset of phase space coordinates, $z_i$,
 $i=1,\cdots,N$, such that
 \be
 \det\left(\frac{\partial \p_i}{\partial z_j}\right)_p\neq 0.
 \ee
 In this case according to the theorem of implicit differentiation
 (or theorem of implicit function), one can in principle, solve
 equations $\p_i(z_i;z'_a)=0$, $i=1,\cdots, N$ to obtain
 $z_i=z_i(z'_a)$, $i=1,\cdots, N$. One can show that the
 set of constraints $\psi_i=z_i-z_i(z'_a)$ which by construction
 are equivalent to $\p_i$'s, are Abelian.
 This can be verified noting that
 \be
 \{\psi_i,\psi_j\}=\{z_i,z_j\}-\{z_i,z_j(z'_a)\}-\{z_i(z'_a),z_j\}+
 \{z_i(z'_a),z_j(z'_a)\},
 \label{b1}\ee
 is independent of $z_i$'s because $\{z^\mu,z^\nu\}=0,\pm1$
 (see Eq.(\ref{symp})). Since
 the left hand side of Eq.(\ref{b1}) vanishes on the constraint
 surface (where $z_i=z_i(z'_a)$), one concludes that it vanishes
 identically and consequently $\psi_i$'s are Abelian \cite{Henbook}.
 Using the chain
 rule of partial differentiation, one can determine explicitly the
 gradient of constraints $\psi_i$'s in terms of the gradient of
 $\p_i$'s, though $\psi_i$'s are implicitly known. This has two
 consequences. Firstly, one can explicitly verify the equivalence
 of gauge transformations generated by $\psi_i$'s and $\p_i$'s.
 Secondly, it determines the homeomorphism  mentioned in section 2
 between the tangent spaces (or gauge groups).
 \par
 The violation of maximality condition has various geometrical
 consequences. For example,
 at any point $p$ where maximality is not satisfied, the
 dimensionality and the norm of the constraint surface is not
 well-defined. Furthermore,
 the tangent space of the constraint surface at $p$ is not homeomorphic to
 the tangent space at the regular points (where
 maximality is satisfied), though tangent spaces at regular points
 are all homeomorphic to each other.
 \par
 In maximality-violated regions, the condition
 $\det(\{\p_i,\q_j\})\neq 0$ on the subsidiary
 constraints (gauge fixing conditions) $\q_i$'s, can not be satisfied
 for any choice of analytic functions $\q_i$'s, because
 \be
 \det(\{\p_i,\q_j\})=\det\left(\frac{\partial \p_i}{\partial
 z^\mu}J^{\mu\nu}\frac{\partial \q_j}{\partial z^\nu}\right),
 \hspace{1cm}i,j=1,\cdots,N,
 \label{FD}
 \ee
 is vanishing if,
 \be
 \mbox{rank}\left(\frac{\partial \p_i}{\partial z^\mu}\right)<N.
 \ee
 In other words, if a set of constraints do not satisfy the
 maximality condition, the Faddeev-Popov determinant (\ref{FD})
 is vanishing for any choice of subsidiary constraints $\q_i$'s.
 \par
 Although maximality is a sufficient condition
 for Abelianization of first class constraints,
 it is not a necessary condition. For example consider
 the constraints of the Friedberg model \cite{Fried}
 $\p_1=p_z$ and $\p_2=xp_y-yp_x$. These
 constraints form a set of Abelian constraint though they do not
 satisfy the maximality condition at the origin.
\par
 To emphasize the topological considerations
 discussed in section 2, let us consider again the SO(3) gauge model.
 One can easily verify that the constraints $L_i$'s do not satisfy the
 maximality condition at any point of phase space. But, since maximality
 is not a necessary condition, using mere this result, one can not conclude
 that $L_i$ are not Abelianizable.
 \subsection{The Space of Gauge Orbits}
 The method of resolution of constraints studied above, considers
 only the equivalence of constraints in the sense that they should define the same
 constraint surface. But it does not verify the equivalence of the corresponding gauge
 groups explicitly. In the following, focusing on the local structure of gauge
 orbits, we obtain a complete description of equivalence of
 constraints and the concept of Abelianizablity.
 \par
 Consider a point $p$ on the constraint surface. The gauge
 transformation generated by the first class constraints $\p_i$,
 $i=1,\cdots,N$ at $p$ can be given as follows,
 \be
 \de_i F=X_i(F)=X^\mu_i\frac{\partial}{\partial z^\mu}F,
 \ee
 where
 \be
 X_i^\mu=\left(\frac{\partial \p_i}{\partial z ^\nu}J^{\nu\mu}\right)_p.
 \label{j}
 \ee
 The vectors $X_i$'s span a subspace of the tangent space $T_p\Phi$
 and determine the direction of gauge
 transformation on the constraint surface at $p$. It is obvious that the maximality
 condition,
 \be
 \mbox{rank} \left(\frac{\partial \p_i}{\partial z^{\mu}}\right)_p=N,
 \label{ind}
 \ee
 is the necessary condition to obtain exactly $N$ independent vector
 $X_i$'s. For further use, note that definition (\ref{j}) implies that
 $X^\mu_i J^{\mu\nu}X_j^\nu=\{\p_i,\p_j\}|_p$ is vanishing identically.
 The gauge fixing conditions $\q_i$'s can be defined as
 functions  which gradient are proportional to $X_i$'s, i.e.
 \be
 \frac{\partial \q_i}{\partial
 z^\mu}=f_iX^\mu_i,\hspace{1cm}\mbox{(no sum over $i$)}
 \label{j1}
 \ee
 where $f_i$ is some function that can be determined by solving the
 condition,
 \be
 \frac{\partial \left(f_iX^\mu_i\right)}{\partial z^\nu}=
 \frac{\partial \left(f_iX^\nu_i\right)}{\partial z^\mu},
  \ee
 which simply means that,
 \be
 \frac{\partial^2 \q_i}{\partial z^\mu\partial
 z^\nu}=\frac{\partial^2 \q_i}{\partial z^\nu\partial z^\mu}.
 \ee
 The gauge fixing conditions $\q_i$'s defined by Eq.(\ref{j1})
 satisfy the following relations:
 \be
 \de_i\q_j=f_iX_i.X_j
 \label{j2}
 \ee
 where $X_i.X_j$ denotes the inner product of two vectors $X_i$ and
 $X_j$ given by the relation $X_i.X_j=X_i^\mu X_j^\mu$.
 \par
 The space of gauge orbits passing through $p$ spanned by the vectors $X^\mu_i$'s,
 can be equivalently spanned by a set of orthonormal
 vectors $\tilde{X}_i$, ($\tilde{X}_i.\tilde{X}_j=\de_{ij}$).
 In appendix 1, we review a well known method to obtain $\tilde{X}_i$'s in terms of
 $X_i$'s. Eqs.(\ref{j}) and (\ref{j1}) can be used to define a new set of
 constraints $\tilde{\p}_i$ and gauge fixing conditions
 $\tilde{\q}_i$'s in terms of $\tilde{X}_i$'s. By construction,
 \be
 \{\tilde{\p}_i,\tilde{\q}_j\}=g_{i}\de_{ij},
 \label{j3}
 \ee
 where $g_{i}$ is some function of phase space. Since we are studying the system
 in an arbitrary small neighborhood of $p$, the function $g_i$ can be estimated as
 a constant that can be absorbed in $\tilde{\q}_i$. Thus Eq.(\ref{j3}) can be rewritten
 in a more interesting form $\{\tilde{\p}_i,\tilde{\q}_j\}=\delta_{ij}$.
 By construction, $\tilde{\p}_i$'s are first class
 constraints equivalent to $\p_i$'s.\footnote{From appendix 1 one verifies that
 $\tilde{X}_i=M_{ij}X_j$
 where $M$ is some invertible matrix. Consequently one can show
 that firstly,
 $\tilde{X}_iJ\tilde{X_j}$ is vanishing on the constraint surface which means
 that $\tilde{\p}_i$'s are first class and secondly, the corresponding gauge groups are
 equivalent}
 Now using the following theorem one can show that the
 set of constraints $\tilde{\p}_i$'s are Abelian.
 \par
 {\bf Theorem.} {\em Given a set of first class constraints $\p_i$, if there exist
 a set of gauge fixing conditions $\q_i$'s such that
 $\{\p_i,\q_j\}=\delta_{ij}$, then $\p_i$'s are Abelian.}
 \par
 {\bf Proof.} Assume that the algebra of constraints is given by
 the relation $\{\p_i,\p_j\}=U_{ij}^{\ \ k}\p_k$.
 Consider the constraints $\p_i$ and $\p_j$ and one
 of gauge fixing conditions, say $\q_k$. Using Jaccobi identity, one can show that,
 \bea
 \left\{\q_k,\{\p_i,\p_j\}\right\}&=&-\left\{\p_j,\{\q_k,\p_i\}\right\}-
 \left\{\p_i,\{\p_j,\q_k\}\right\}\nn\\
 &=&\{\p_j,\de_{ki}\}-\{\p_i,\de_{jk}\},
 \eea
 which is vanishing identically. Therefore,
 \bea
 0&=&\left\{\q_k,\{\p_i,\p_j\}\right\}\nn\\
 &=&\{\q_k,U_{ij}^{\ \ k'}\p_{k'}\}\nn\\
 &=&U_{ij}^{\ \ k}-U_{ij}^{\ \ kk'}\p_{k'},
 \eea
 where $U_{ij}^{\ \ kk'}=-\{\q_k,U_{ij}^{\ \ k'}\}$.
 Consequently the algebra of constraints is given as follows,
 \be
 \{\p_i,\p_j\}=U_{ij}^{\ \ kk'}\p_k\p_{k'}.
 \ee
 By repeating the above calculation, arbitrary number of $\p_i$
 can appear on the right hand side of the above relation. To obtain a
 meaningful algebra, this chain of multiplying constraints should
 terminate somewhere which means that the right hand side of the
 above equation is vanishing identically. In other words, $\p_i$'s
 are Abelian.
 \par
 Summarizing our results, if at some point $p$ on the surface of $N$ first class
 constraints, the maximality condition is satisfied then the space of
 gauge orbits at that point looks like ${\cal R}^N$ (spanned by $N$ orthogonal
 vectors $\tilde{X}_i$'s). Furthermore
 there exist an equivalent set of Abelian constraints which can be
 recognized as generators of translation along orthogonal
 directions of ${\cal R}^N$. Consequently, if we are given a finite set
 of first class constraints, such that the corresponding space of
 gauge orbits looks like, say, a sphere, they  satisfy the
 maximality condition nowhere on the constraint surface. Therefore
 for any choice of subsidiary constraints, the Faddeev-Popov
 determinant is vanishing. This result can also be verified noting
 that the Faddeev-Popov determinant is, in general, proportional to
 $\det(X_i.X_j)_{N\times N}$ which can be non-vanishing only if
 maximality condition is satisfied and $X_i$'s are exactly
 $N$ independent vectors (See Eq.(\ref{ind})).
 \par
 Finally let us briefly review the Abelianization method given in
 ref.\cite{Abelian}. There, using the Cauchy-Kowalevski theorem
 \cite{John}, it is concluded that for any analytic constraint $\p$,
 the partial differential equation $\{\q,\p\}=1$ has at least one
 solution for $\q$ that can be uniquely determined by the boundary
 conditions. $\q$ is used to define a projection map to the constraint surface
 $\p=0$. It is shown that after projecting the remaining
 constraints to the surface $\p=0$, one finds a new set of constraints
 equivalent to the
 original set with an interesting property: the Poisson brackets of all mapped constraints
 with $\p$ is vanishing identically. It is shown that using similar projection operators one
 can map all constraints to the surface of each other consistently,
 which results in an equivalent Abelian set of constraints. In
 addition it shown that one can obtain a set of subsidiary
 constraint such that $\{\p_i,\q_j\}=\de_{ij}$.\footnote{It is
 straightforward to show that, obtaining the projected constraints is equivalent to
 calculating the orthogonal vectors $\tilde{X}_i$'s in terms of the original
 $X_i$'s.}
 Obviously, the domain of validity of this method is  determined by the domain of validity
 Cauchy-Kowalevski theorem. As one anticipates, this theorem is valid as far as the
 maximality condition is satisfied. See ref.\cite{John} for
 details.
 \par
 The result of this section can be used to prove the following
 theorem on second class constraints:
 \par
 {\bf Theorem} {\em If a given set of second class constraints can be
 considered as the union of first class constraints and the
 corresponding gauge fixing conditions, then the subset of first
 class constraints is Abelianizable.}
 \par
 To prove this theorem note that by definition a set of second
 class constraints $\psi_I$'s satisfy the relation
 $\det(\{\psi_I,\psi_J\})|_\psi\neq 0$. If $\psi_I$'s are a combination of some first
 class constraints $\p_i$ (namely $\{\p_i,\p_j\}|_\psi=0$) and gauge fixing conditions
 $\q_i$, the definition of second class constraints gives,
 \be
 \det\left(\begin{array}{cc}
 0&A\\
 -A&\{\q_i,\q_j\}
 \end{array}\right)\neq 0
 \ee
 where $A_{ij}=\{\p_i,\q_j\}$. Thus $\det A\neq 0$ and the
 maximality condition is satisfied.  For more details see
 ref.\cite{Second} and references therein.
 \section{Examples}
 In this section we consider the  simplest examples of
 Abelianizable and non-Abelianizable constraints. The non-Abelianizable constraints that
 we study here are the non-Abelian constraints of $SO(3)$ gauge model. We
 explicitly show that the Faddeev-Popov determinant is vanishing
 for any choice of gauge fixing conditions.
 \par
 {\bf 1. Abelianizable Constraints.} Consider a system given by the
 following constraints,
 \be
 \p_1=p_x,\hspace{1cm}\p_2=p_y-e^xp_y,
 \ee
 which satisfy the algebra, $\{\p_1,\p_2\}=e^x\p_1$. Obviously, this set of constraints is
 equivalent  to the Abelian set $\tilde{\p}_1=p_x$ and $\tilde{\p}_2=p_y$.
 It is interesting to obtain these Abelian constraints using the
 method of subsection 3.2. First note that
 \be
 X_1=\left(\begin{array}{c}1\\0\\0\\0\end{array}\right),\hspace{1cm}
 X_2=\left(\begin{array}{c}-e^x\\1\\0\\0\end{array}\right),
 \ee
 where, for example, $X_1=J.\nabla\p_1|_{\p_1,\p_2}$, in which
 \be
 \nabla\p_1=\left(\begin{array}{c}\frac{\partial}{\partial x}\\
 \frac{\partial}{\partial y}\\
 \frac{\partial}{\partial p_x}\\
 \frac{\partial}{\partial p_y}\end{array}\right)\p_1,
 \ee
 Using the method of Appendix 1, one obtains,
 \be
 {\tilde X}_1=\left(\begin{array}{c}1\\0\\0\\0\end{array}\right),\hspace{1cm}
 {\tilde X}_2=\left(\begin{array}{c}0\\1\\0\\0\end{array}\right),
 \ee
 and consequently
 \be
 \nabla{\tilde \p}_1=-J.\tilde{X}_1=\left(\begin{array}{c}0\\0\\1\\0\end{array}\right),
 \hspace{1cm}
 \nabla{\tilde \p}_2=-J.\tilde{X}_2=\left(\begin{array}{c}0\\0\\0\\1\end{array}\right),
 \ee
 which gives, $\tilde{\p}_1=p_x$ and $\tilde{\p}_2=p_y$.
 \par
 {\bf 2. Non-Abelianizable Constraints.} Consider the constraints of
 $SO(3)$ gauge model,  $L_i=\epsilon_{ijk}x_jp_k$, $i=1,2,3$.
 Assume three arbitrary subsidiary constraints $\q_i$'s. Since
 $L_i$'s are non-Abelianizable, the Faddeev-Popov determinant
 $\det(\{\q_i,L_j\})$ is vanishing as can be verified as follows.
 Using the equality,
 \be
 \det\left(\begin{array}{ccc}a_{11}&a_{12}&a_{13}\\a_{21}&a_{22}&a_{23}\\
 a_{31}&a_{32}&a_{33}\end{array}\right)=\epsilon_{ijk}a_{1i}a_{2j}a_{3k},
 \ee
 one finds that
 \bea
 \det(\{\q_i,L_j\})&=&\epsilon_{ijk}\{\q_1,L_i\}\{\q_2,L_j\}\{\q_3,L_k\}\nn\\
 &=&-\epsilon_{ijk}\epsilon_{ia_1b_1}\epsilon_{ja_2b_2}\epsilon_{ka_3b_3}\prod_{c=1}^3
 \left(\frac{\partial \q_c}{\partial
 x_{a_c}}x_{b_c}+\frac{\partial \q_c}{\partial
 p_{a_c}}p_{b_c}\right),
 \eea
 Two generic terms in the above sum can be distinguished:
 \be
 P=\epsilon_{ijk}\epsilon_{iaa'}\epsilon_{jbb'}\epsilon_{kcc'}
 \left(\frac{\partial \q_1}{\partial x_a}x_{a'}\right)
 \left(\frac{\partial \q_2}{\partial x_b}x_{b'}\right)
 \left(\frac{\partial \q_3}{\partial x_c}x_{c'}\right),
 \ee
 and
 \be
 Q=\epsilon_{ijk}\epsilon_{iaa'}\epsilon_{jbb'}\epsilon_{kcc'}
 \left(\frac{\partial \q_1}{\partial x_a}x_{a'}\right)
 \left(\frac{\partial \q_2}{\partial x_b}x_{b'}\right)
 \left(\frac{\partial \q_3}{\partial p_c}p_{c'}\right),
 \ee
 To calculate $P$, one realizes three generic terms:
 \bea
 P_1&=& \left(\frac{\partial \q_1}{\partial x}\right)
 \left(\frac{\partial \q_2}{\partial x}\right)
 \left(\frac{\partial \q_3}{\partial x}\right)
 \epsilon_{ijk}\epsilon_{i1a}\epsilon_{j1b}\epsilon_{k1c}(x_ax_bx_c),\nn\\
 P_2&=& \left(\frac{\partial \q_1}{\partial x}\right)
 \left(\frac{\partial \q_2}{\partial x}\right)
 \left(\frac{\partial \q_3}{\partial y}\right)
 \epsilon_{ijk}\epsilon_{i1a}\epsilon_{j1b}\epsilon_{k2c}(x_ax_bx_c),\nn\\
 P_3&=& \left(\frac{\partial \q_1}{\partial x}\right)
 \left(\frac{\partial \q_2}{\partial y}\right)
 \left(\frac{\partial \q_3}{\partial z}\right)
 \epsilon_{ijk}\epsilon_{i1a}\epsilon_{j2b}\epsilon_{k3c}(x_ax_bx_c).
 \label{calP}
 \eea
 $P_1=0$ because here, $(i,j,k)\in\{2,3\}$ and consequently $\epsilon_{ijk}$ is
 vanishing. $P_2=0$ because $\epsilon_{i1a}\epsilon_{j2_b}x_ax_b$
 is symmetric under $a\leftrightarrow b$ and consequently under $i\leftrightarrow j$,
 though $\epsilon_{ijk}=-\epsilon_{jik}$. In addition,
 $P_3=-yzx+zxy=0$ (The first term corresponds to $a=2$
 and the second terms corresponds to $a=3$ in Eq.(\ref{calP})).
 \par
 $Q$ is the sum of four generic terms:
 \bea
 Q_1&=&\frac{\partial \q_1}{\partial x}\frac{\partial \q_2}{\partial x}
 \frac{\partial \q_3}{\partial p_x}\epsilon_{ijk}\epsilon_{i1a'}\epsilon_{j1b'}
 \epsilon_{k1c'}x_{a'}x_{b'}p_{c'},\nn\\
 Q_2&=&\frac{\partial \q_1}{\partial x}\frac{\partial \q_2}{\partial x}
 \frac{\partial \q_3}{\partial p_y}\epsilon_{ijk}\epsilon_{i1a'}\epsilon_{j1b'}
 \epsilon_{k2c'}x_{a'}x_{b'}p_{c'},\nn\\
 Q_3&=&\frac{\partial \q_1}{\partial x}\frac{\partial \q_2}{\partial y}
 \frac{\partial \q_3}{\partial p_x}\epsilon_{ijk}\epsilon_{i1a'}\epsilon_{j2b'}
 \epsilon_{k1c'}x_{a'}x_{b'}p_{c'},\nn\\
 Q_4&=&\frac{\partial \q_1}{\partial x}\frac{\partial \q_2}{\partial y}
 \frac{\partial \q_3}{\partial p_z}\epsilon_{ijk}\epsilon_{i1a'}\epsilon_{j2b'}
 \epsilon_{k3c'}x_{a'}x_{b'}p_{c'}.
 \eea
 $Q_1$ and $Q_2$ are vanishing because under $i\leftrightarrow j$,
 $\epsilon_{i1a'}\epsilon_{j1b'}x_{a'}x_{b'}$ is symmetric but
 $\epsilon_{ijk}$ is antisymmetric. Using the identity,
 $\epsilon_{ijk}\epsilon_{j2b'}=-\de_{i2}\de_{kb'}+\de_{ib'}\de_{k2}$,
 one can  show that $Q_3$ and $Q_4$ are some combinations of $L_i$'s.
 Therefore $Q$ is also vanishing on the constraint surface.
 \section{Conclusion}
 We found that first class constraints can be classified as Abelianizable
 and non-Abelianizable constraints. These classes are identified by
 topological invariants (e.g. compactness) of
 the corresponding gauge groups. The topology of a gauge
 group is uniquely determined by the structure coefficients of the
 gauge generators' algebra which are simply the structure
 functions of  the constraint algebra calculated at some particular
 point of phase space. Since maximality is the necessary condition
 on a given set of first class constraints
 for existence of a proper set of gauge fixing conditions, i.e. a
 set of subsidiary constraints such that the Faddeev-Popov
 determinant is non-vanishing, we concluded that
 \begin{enumerate}
 \item{These constraints are Abelianizable if there exist a set of subsidiary constraints
 such that Faddeev-Popov determinant is non-vanishing.}
 \item{If these constraints are non-Abelianizable then
 Faddeev-Popov determinant is vanishing for any choice of gauge
 fixing conditions.}
 \end{enumerate}
 We studied the $SO(3)$ gauge invariant model as an example.
 \par
 Using the first result mentioned above, we found that if a set of
 second class constraints is considered as the union of first
 class constraints and the corresponding gauge fixing conditions,
 then the subset of first class constraints is Abelianizable.
  \section{Appendix 1}
 Here we briefly review the method of obtaining $N$ orthogonal vectors ${\vec u}_i$ in
 terms of a given set of $N$ linearly independent vectors ${\vec v}_i$. This is a
 well know method that can be found in elementary text books
 in mathematics. Assuming that ${\vec v}_i.{\vec v}_i=1$, the set of orthogonal
 vectors ${\vec u}_i$'s, up to some normalization constants can be obtained as follows:
 \bea
 {\vec u}_1&=&{\vec v}_1,\nn\\
 {\vec u}_2&=&{\vec v}_2-({\vec u}_1.{\vec v}_2){\vec u}_1,\nn\\
 {\vec u}_{n+1}&=&v_n-\sum_{i=1}^n(u_i.{\vec
 v}_n)u_i,\hspace{1cm}n=2,\cdots,N.
 \eea
 \section{Appendix 2}
 In this appendix we study the algebra of $SU(N)$ Yang-Mills
 theory. It is well know that this model possesses an infinite
 set of non-Abelian constraints,
 \be
 \p_a({\vec x})=\partial_i\Pi_i^a-gf_{abc}A^b_i\Pi^c_i,
 \ee
 where $f_{abc}$'s are the structure coefficients of $SU(N)$ algebra and
 $\Pi^a_i$ is the momentum field conjugate to gauge field
 $A^a_i$, i.e. $\{A^a_i({\vec x}),\Pi^b_j({\vec
 y})\}=\de^{ab}\de_{ij}\de^D({\vec x}-{\vec y})$.
 These constraints satisfy the following algebra,
 \be
 \{{\p_a}^{g_1},{\p_b}^{g_2}\}=gf_{abc}{\p_c}^{g_1g_2},
 \ee
 in which
 $g_1({\vec x})$ and $g_2({\vec x})$ are some smooth functions and
 ${\p_a}^g=\int d^3x g(x)\p_a(x)$ \cite{Hen,Henbook}.
 Assuming $g_i=e^{-ip_ix}$ where the momentum space is a
 compactified lattice, which corresponds to  lattice
 gauge theory on tori, we obtain a finite set of constraints satisfying
 the following {\bf closed} algebra:
 \be
 \{\p^m_a,\p^n_b\}=gf_{abc}\de_{m+n-p}\p^p_c.
 \ee
 $\p^m_a$'s are non-Abelianizable and consequently the
 corresponding Faddeev-Popov determinant is vanishing. It is
 important to examine any possible relation between this result
 and the appearance of Gribov copies in ordinary $SU(N)$ Yang-Mills theory.
 \section*{Acknowledgement} The financial support of Isfahan
 University of Technology (IUT) is acknowledged.


\begin{thebibliography}{99}
\bibitem{Dirac} P. A. M. Dirac, Can. J. Math. {\bf 2}, (1950) 129 ;
Proc. R. Soc. London Ser. A {\bf 246}, (1958) 326; {\it "Lectures
on Quantum Mechanics"} New York: Yeshiva University Press, 1964,
\bibitem{Pons} C. Batlle, J. Gomis, X. Gracia and J. M. Pons, J.
Math. Phys. {\bf 30} (6), (1989) 1345; J. M. Pons and J. A.
Garcia, Int. J. Mod. Phys. A {\bf 15} (2000) 4681, hep-th/9908151.
\bibitem{Hen} M. Henneaux, Phys. Rep. {\bf 126}, (1985) 1;
\bibitem {Henbook} M. Henneaux and C. Teitelboim {\it "Quantization of
Gauge System"} Princeton University Press, Princeton, New Jersey,
1992.
\bibitem{Gold} J. Goldberg, E. T. Newman and C. Rovelli, J. Math. Phys.
{\bf 32}, (1991) 2739.
\bibitem{Berg} P. G. Bergman and I. Goldberg, Phys. Rev. {\bf 98},
(1955) 531; P. G.
Bergman {\it ibid}. {\bf 98}, (1955) 544.
\bibitem{Gog} S. A. Gogilidze, A. M. Khvedelidze and V. N. Pervushin,
J. Math. Phys. {\bf 37}, (1996) 1760, hep-th/9504153.
\bibitem{Abelian} F. Loran, Phys. Lett. {\bf B547}, (2002) 63,
hep-th/0209180.
\bibitem {Second} F. Loran, Phys. Lett. {\bf B554}, (2003) 207,
hep-th/0212341.
\bibitem{Battle} C. Battle, J. Gomis, J.M. Pons and N. Roman-Roy,
J. Math. Phys. {\bf 27} (12) (1986) 2953.
\bibitem{Govbook}
J. Govaerts, {\it "Hamiltonian Quantisation and Constrained
Dynamics"}, Leuven Notes in Theoretical and Mathematical Physics,
Leuven University Press, 1991.
\bibitem{Munk} J. R. Munkres, {\em "Topology, A First Course"},
Prentice-Hall, Inc, Englewood Cliffs, New~Jersey, 1975.
\bibitem{Govpap} J. Govaerts and J. R. Klauder,
Annals Phys. {\bf 274} (1999), 251, hep-th/9809119.
\bibitem{Fried}  R. Friedberg, T. D Lee, Y. Pang and
H. C. Ren, Ann. Phys. {\bf 246}, (1996)
381.
\bibitem{John} F. John, "{\it Partial Differential Equations}", vol. 1, Fourth Edition,
Springer-Verlag, New York Inc., 1981.
\end{thebibliography}
\end{document}